\documentclass[journal=jacsat,manuscript=article]{achemso}
\setkeys{acs}{maxauthors=20}
\setkeys{acs}{etalmode=truncate}

\usepackage{graphicx}
\usepackage{amsmath,amssymb,latexsym,xspace}
\usepackage{dcolumn}
\usepackage{bm}
\usepackage{subfigure}

\author{Kevin M. Conley}
\affiliation[Aalto]
{Department of Chemistry and Materials Science, Aalto University School of Chemical Engineering, P.O. Box 16100, FI-00076 Aalto, Finland}
\author{Caterina Cocchi}
\affiliation[Oldenburg]
{Institut für Physik, Carl von Ossietzky Universität Oldenburg, 26129 Oldenburg, Germany}
\alsoaffiliation[berlin]{Institut für Physik und IRIS Adlershof, Humboldt-Universität zu Berlin, Zum Großen Windkanal 2, 12489 Berlin, Germany}
\author{Tapio Ala-Nissila}
\email{tapio.ala-nissila@aalto.fi}
\affiliation[Aalto2]
{QTF Centre of Excellence, Department of Applied Physics, Aalto University School of Science, P.O. Box 11100, FI-00076 Aalto, Finland}
\alsoaffiliation[Lboro]
{Interdisciplinary Centre for Mathematical Modelling, Department of Mathematical Sciences, Loughborough University, Loughborough, Leicestershire LE11 3TU, United Kingdom}

\title{Formation of Near-IR Excitons in Low Dimensional CuSbS$_2$}

\begin{document}

\clearpage

\begin{abstract}\normalsize
    The electronic and optical properties of low-dimensional semiconductors are typically quite different from those of their bulk counterparts. Yet, the optical gap of two-dimensional copper antimony disulfide (CuSbS$_2$) does not dramatically change with decreasing thickness of the material. The absorption onset remains at about 1.5 eV in the monolayer, bilayer, and bulk materials. Using density functional theory and many-body perturbation theory, we rationalize this behavior through the interplay of quantum confinement, electron-hole interactions, and the formation of surface states. Specifically, the spatial confinement in thin layers induces strongly bound optical transitions in the near-infrared region. Our results explain the optical properties in copper antimony disulfide platelets of varying thickness and set these materials as potential candidates for novel photovoltaic devices and near-infrared sensors. 
\end{abstract}

\clearpage

\section{Introduction}

Thin-film materials for solar photovoltaics are challenging traditional wafer silicon in the renewable energy sector~\cite{walsh2012kesterite, mitzi2013prospects, pal2019current, ramanujam2020flexible, peter2011towards}. Scarcity and competing industrial demands for the constituent elements of CdTe and copper indium gallium disulfide/diselenide devices have driven the search for alternative systems with comparable performance~\cite{peter2011towards, fthenakis2009life, peccerillo2018copper}. Among them, CuSbS$_2$ is a promising material formed by earth abundant elements and characterized by a large absorption coefficient at relevant frequencies for solar light harvesting~\cite{kumar2014cu, de2017characterization}. The optical gap of bulk CuSbS$_2$ ranges between 1.3 and 1.5 eV~\cite{rodriguez2001cusbs2, kyono2005crystal, zhou2009solvothermal, ikeda2018structural, peccerillo2018copper}. Theoretical studies based on quantum density functional theory (DFT) calculations~\cite{hohe-kohn64pr,kohn-sham65pr} have shown that in this material, the fundamental gap is indirect, like in bulk silicon, and that the direct gap appears at a marginally higher energy of 1.73 eV~\cite{temple2012geometry, dufton2012structural, birkett2016optical, gassoumi2017investigation, gupta2019thermodynamic}. 

Recently, single crystal CuSbS$_2$ platelets of varying thickness have been experimentally prepared using hot injection method~\cite{moosakhani2018platelet,moosakhani2018solution}. This bottom-up approach enables the synthesis of nanomaterials with well-defined size and homogeneous shape down to monolayer thickness~\cite{moosakhani2018solution}.
Interestingly, in CuSbS$_2$, quantum confinement effects accompanying the nanostructuring process do not lead to a sizeable increase of the optical gap, as typically observed in low-dimensional semiconductors and insulators in comparison with their bulk counterparts~\cite{novoselov2004electric,butler2013progress,liu2017two,turkowski2017time,mounet2018two}.
For example, in transition metal dichalcogenides, spatial confinement increases absorption and photoluminescence~\cite{mak2010atomically, jariwala2014emerging, voshell2018review, mueller2018exciton, lau2019electronic, edalati2020mo, liang2021defect} or even induces superconductivity~\cite{choi2017recent}.
Enhancement of optical activities have been also observed in hexagonal boron nitride~\cite{cassabois2016hexagonal, paleari2018excitons, wickramaratne2018monolayer} and heterostructures thereof~\cite{aggoune2017enhanced,bawari2019enhanced}, as well as in the more recently produced silicene~\cite{li2013intrinsic, hui2015silicene, houssa2015silicene, chowdhury2016theoretical}.

Such peculiar behavior of CuSbS$_2$ has been already explored in a DFT study~\cite{ramasamy2014mono}, which, however, has not been able to convincingly rationalize the change of the optical gap of thin platelets (1 -- 5~nm) of this material compared to the bulk. 
Understanding the fundamental mechanism that is responsible for this counterintuitive behavior is essential not only to fully exploit the potential of CuSbS$_2$ in optoelectronic and photovoltaic applications.
Most importantly, it is mandatory to gain further insight into exotic behaviors of quantum-confined semiconductors.
  

\begin{figure}[hbt!]
    \includegraphics[width=\textwidth]{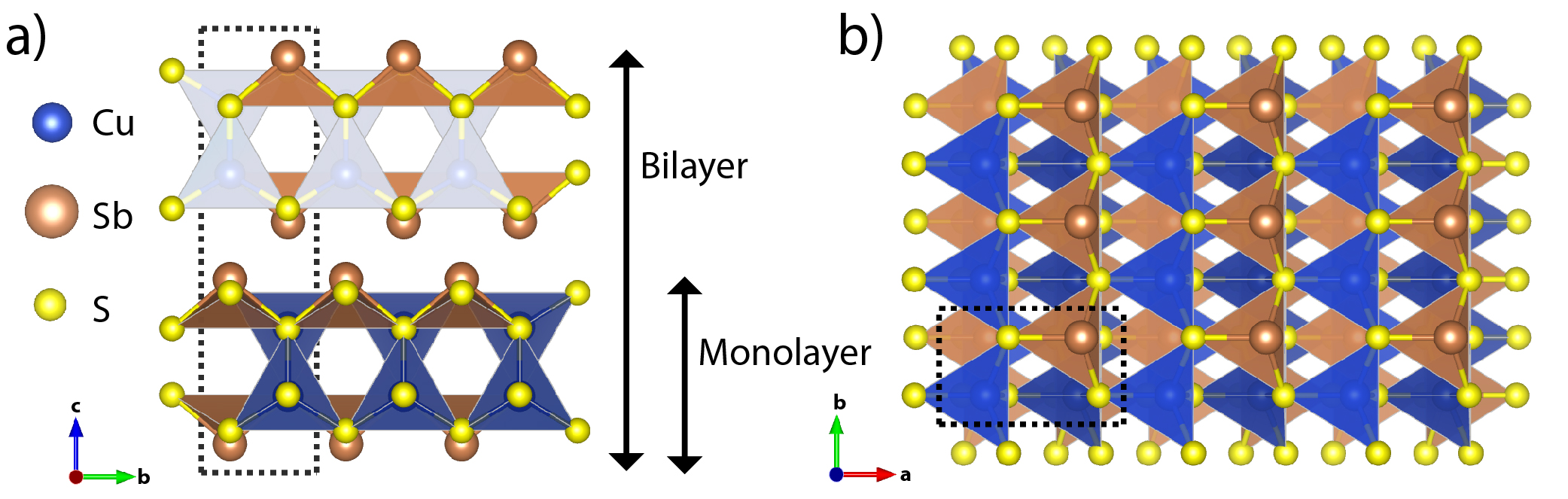}
    \caption{a) Layered and b) bulk CuSbS$_2$ with Cu, Sb, and S atoms shown as blue, brown, and yellow spheres.}
    \label{fig:gap_thickness}
\end{figure}

In this work, based on DFT and many-body perturbation theory (MBPT)~\cite{onid+02rmp}, we rationalize why the optical gap of two dimensional CuSbS$_2$ platelets does not increase compared to their bulk counterparts.
From the analysis of the band structure, the density of states, the charge-density distribution and, finally, the dielectric and optical response of this material in the monolayer, bilayer, and bulk form, we ascribe this behavior to the concurrent contribution of localized surface states, that are formed in the nanoconfined systems, and strongly bound excitons that appear therein. 

\section{Methods}
All calculations presented in this work are performed with the Vienna \textit{ab-initio} simulation package (VASP)~\cite{kresse1996efficiency} utilizing the projector-augmented wave (PAW) method~\cite{blochl1994projector, kresse1999ultrasoft}. The electronic structure was calculated using DFT with the Heyd-Scuseria-Ernzerhof (HSE06) hybrid functional~\cite{heyd2003hybrid}. A $4~\times~6~\times~2$ $k$-point mesh and 370 eV cutoff is used and the structures were assumed to be relaxed when the forces on each atom were smaller than 10~meV~\AA{}$^{-1}$. The mono- and bilayer structures are simulated including in the unit cell 9~\AA{} of vacuum along the (001) surface in order to prevent unphysical interactions between the replicas. We checked this value by ensuring the band gap remained constant with additional vacuum, and by checking that the total charge density decays to less than 1$\times10^{-6}~a_0^{-3}$ in the non-periodic direction. 

The quasiparticle band structures are calculated with a single-shot G$_0$W$_0$ approach applied on top of the HSE06 electronic structure. The GW band structure converged with 600 bands or 320 bands in the case of the monolayer. The optical properties were calculated by solving the Bethe--Salpeter equation (BSE)~\cite{salpeter1951relativistic} and include contributions beyond the Tamm-Dancoff approximation~\cite{sander2015beyond}. The 30 highest valence and the 60 lowest conduction bands are considered in the bilayer and bulk and the 28 highest valence and the 60 lowest conduction bands are considered in the monolayer. Spurious contributions arising from the vacuum in the layered structures are excluded by rescaling the dielectric permittivity with the effective volume of the layer. 

\section{Results and Discussion}
The optimized lattice constants and atomic positions obtained for bulk CuSbS$_2$ (Figure~\ref{fig:gap_thickness}) are consistent with previous findings in the literature obtained on similar footing~\cite{temple2012geometry, dufton2012structural} (see Table S1 in the Supporting Information). CuSbS$_2$ is an orthorhombic crystal with space group \textit{Pnma}. The space groups of the bilayer and monolayer are \textit{P2$_1/m$} and \textit{Pm}, respectively. In the layered crystals, no surface reconstruction is found, although the atoms are bound closer together normal to the (001) surface. Specifically, the material volume in the monolayer shrinks by 4.2\% compared to the bulk, as an effect of the quantum confinement along the $c$ axis (see Figure~\ref{fig:gap_thickness}a), in agreement with experimental findings on powder CuSbS$_2$ platelets~\cite{moosakhani2018solution}. In the bilayer, this reduction is only 0.8\%. 

\begin{figure}[hbt!]
    \includegraphics{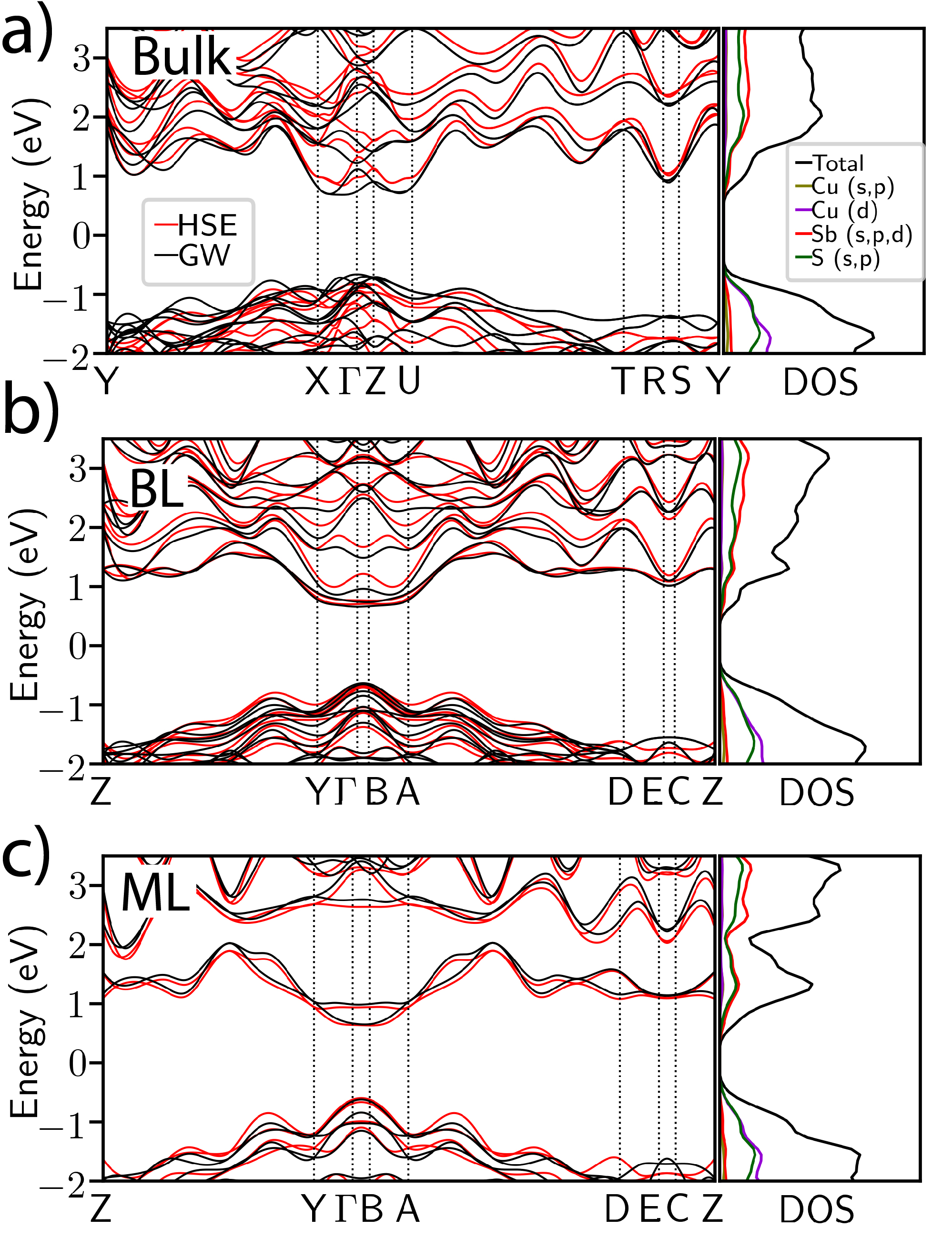}
    \caption{GW (black lines) and HSE06 band structure (red lines) of a) bulk, b) bilayer (BL), and c) monolayer (ML) CuSbS$_2$. The density of states and atom-projected contributions calculated using HSE06 are shown adjacent to the band structures.}
    \label{fig:gw_bands}
\end{figure}

From the analysis of the electronic properties, we find for bulk CuSbS$_2$ an indirect fundamental gap of 1.67 eV, as predicted from DFT using HSE06 (see Figure~\ref{fig:gw_bands}a), in agreement with previous calculations~\cite{ramasamy2014mono,temple2012geometry,dufton2012structural}. Despite quantum confinement, both monolayer and bilayer CuSbS$_2$ exhibit smaller band gaps than the bulk, with the former featuring the narrowest gap. 
The inclusion of the quasi-particle correction via the perturbative G$_0$W$_0$ approach applied on top of the HSE06 electronic structure leads to a reduction of the band gap in both bulk and bilayer, and to an increase in the monolayer.
The nature of the gap (indirect in the bulk and direct in the low-dimensional sheets) is unaffected by the inclusion of the self-energy contribution (see Table~\ref{tab:optical}). 


The crossover from indirect- to direct-band-gap material going from bulk to mono- and bilayer is ubiquitous in layered semiconductors~\cite{mak2010atomically,tongay2012thermally,edalati2020mo,mudd2016direct}. In order to understand this behavior in CuSbS$_2$, we analyze the band dispersion shown in Figure~\ref{fig:gw_bands}. The Brillouin zone of each system is shown in Figure S1. In the bulk material, the valence band maximum of bulk CuSbS$_2$ lies between the $\Gamma$ and $Z$ points and the conduction band minimum lies between the $\Gamma$ and $X$ points. The smallest, direct gap lies between the $\Gamma$ and $X$ points. Interestingly, in the layered materials, the highest valence band and the lowest conduction bands flatten along the $B$ and $Y$ directions. This change in the band dispersion is responsible for the crossover to a direct band gap. 
The variation in the fundamental band gap in increasingly confined layers is attributed to the reduced band dispersion accompanied by the formation of a surface state at the Sb-terminated (001) surface. The states have Cu $d$ and S $sp$ character at the top of the valence band and S $sp$ and Sb $spd$-character at the bottom of the conduction band (Figure~\ref{fig:gw_bands}). There is no significant change in the character of the states in the layered materials. The partial DOS of bulk CuSbS$_2$ is consistent with previous DFT studies in bulk CuSbS$_2$~\cite{gupta2019thermodynamic}. 

\begin{table}[hbt!]
\begin{center}
\caption{Band gaps and optical transitions of bulk and layered CuSbS$_2$.\label{tab:optical}}
 \begin{tabular}{|l | r r r r|} 
 \firsthline
   &\multicolumn{1}{l}{Indirect band}&\multicolumn{1}{l}{Direct band}&  \multicolumn{1}{l}{$E_{\textrm{optical}}$ (eV)}&\multicolumn{1}{l|}{$E_\textrm{b}$ (eV)} \\
   &\multicolumn{1}{l}{gap (eV)}&\multicolumn{1}{l}{gap (eV)}&\multicolumn{1}{l}{} & \multicolumn{1}{l|}{}\\ [0.5ex] 
 \hline
 \textbf{Bulk} & & & &\\
 This work, MBPT    & 1.35 & 1.40 & 1.20 & 0.20 \\ 
 This work, HSE06 & 1.67 & 1.71 & & \\ 
 HSE06~\cite{temple2012geometry} & 1.68 & 1.73 & &\\
 HSE06~\cite{dufton2012structural}& 1.69 & & &\multicolumn{1}{l|}{} \\  
 \hline
 \textbf{Layered}  & & & & \\
 This work, MBPT BL & & 1.30  & 0.95   & 0.35  \\
 This work, MBPT ML & & 1.30  & 0.76   & 0.54  \\
 This work, HSE06 BL & & 1.38 & &  \\
 This work, HSE06 ML & & 1.25 & &  \\
 \hline
 \textbf{Experimental}  & & & & \\
 Platelets (25-50 nm thick)~\cite{moosakhani2018solution} & 1.51$^a$ & 1.57$^a$ & 0.92$^b$ & 0.65$^c$ \\
 Platelets (1-5 nm thick)~\cite{moosakhani2018solution} & 1.44$^a$ & 1.52$^a$ & 1.04$^b$ & 0.48$^c$ \\
 \lasthline
\end{tabular}
\begin{flushleft}
$^a$ Optical gap measured using diffuse reflectance spectroscopy.\\
$^b$ Energies less than 0.85 eV were beyond the experimental range.\\
$^c$ The experimental binding energy, $E_\textrm{b}$, was obtained by subtracting the optical transition energy from the direct optical gap.
\end{flushleft}
\end{center}
\end{table}

\begin{figure}[hbt!]
    \includegraphics[width=\textwidth]{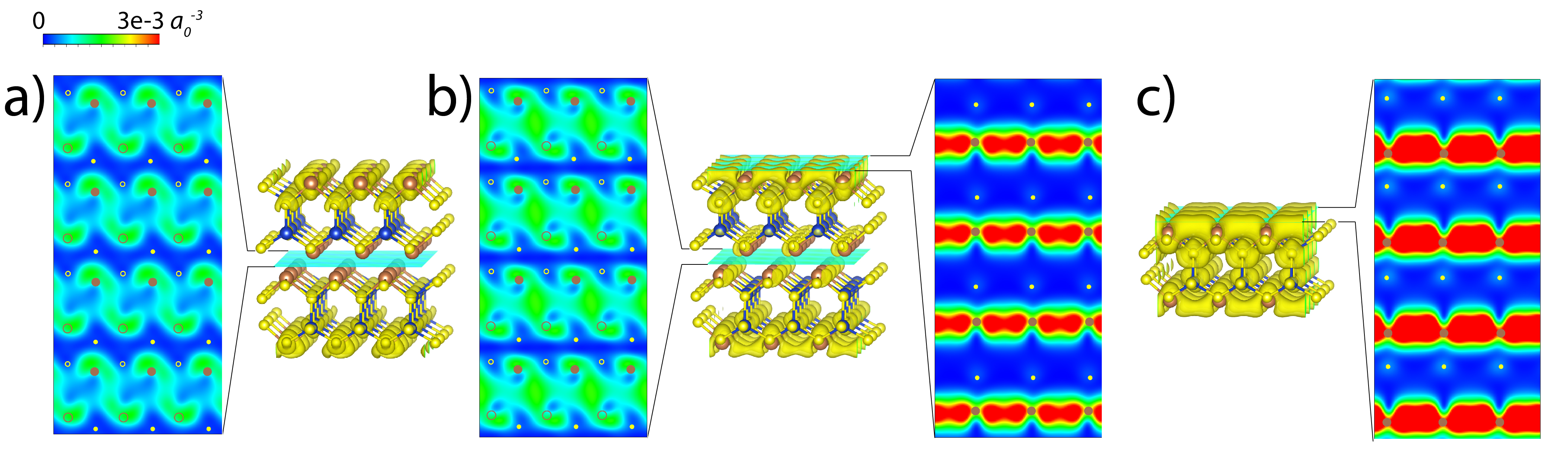}\\ 
    \caption{Spatial distribution of the probability density associated to the conduction band minimum of a) bulk, b) bilayer, and c) monolayer CuSbS$_2$ calculated from DFT. The displayed isosurface value is 0.003 $a_0^{-3}$. The positions of atoms above (below) the considered plane are marked by open (closed) circles, brown for Sb and yellow for S.} 
    \label{fig:surface_states}
\end{figure}

To confirm the presence of surface states in the Sb-terminated dangling surface of mono- and bilayer CuSbS$_2$, we inspect the electron density at the conduction band minimum. The results plotted along a two-dimensional slice in the (001) plane (see Figure~\ref{fig:surface_states}) confirm that the electron density accumulates at the surface of the low-dimensional structures. The electron density between the layers in the bilayer is similar to the interlayer electron density in the bulk material. The electron density spills out of the monolayer more than the bilayer (Figure~\ref{fig:surface_states}c). This results in electron density accumulation between the adjacent surface Sb atoms. 

\begin{figure}[hbt!]
\includegraphics{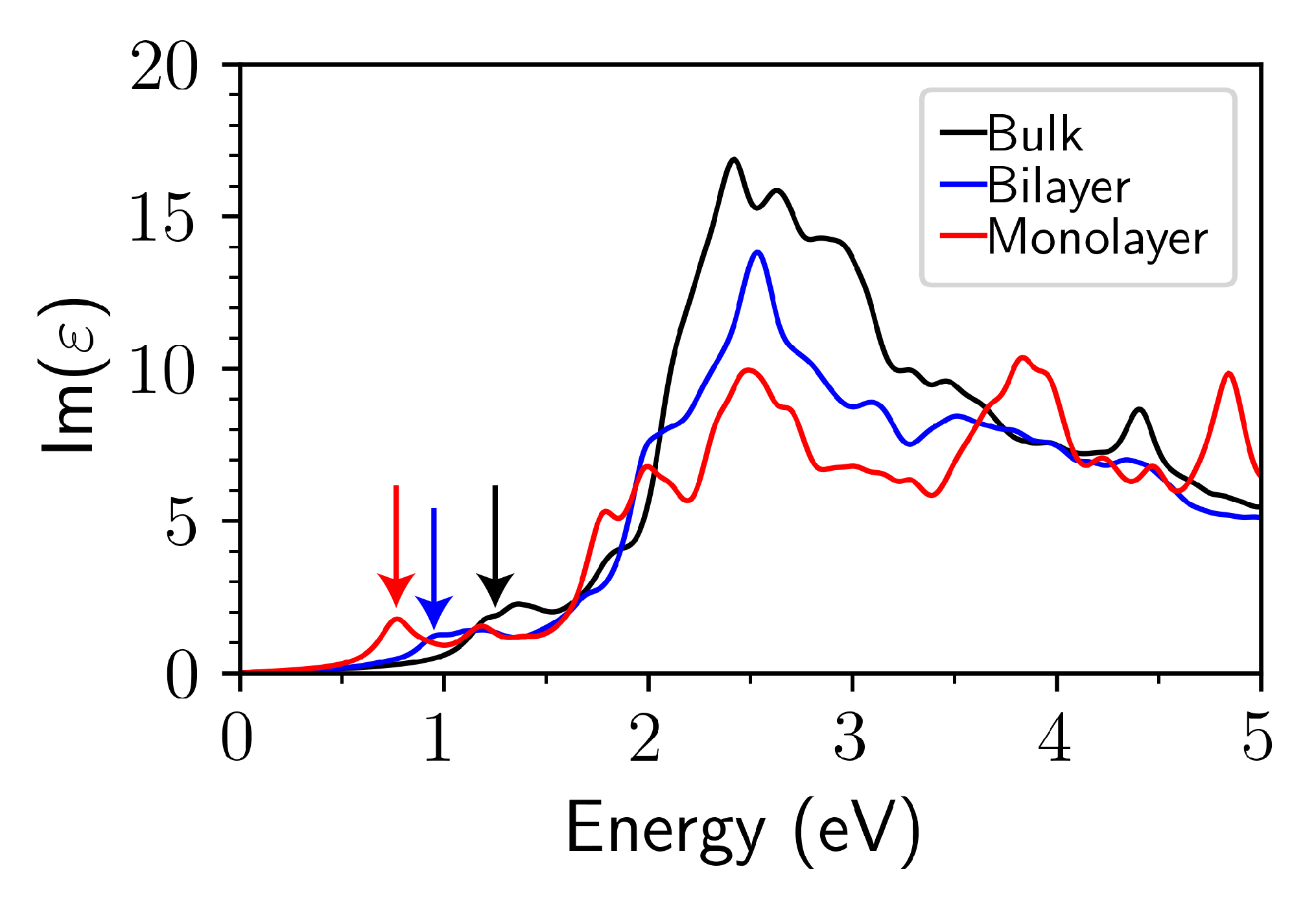}
    \caption{Imaginary component of the dielectric permittivity of bulk, bilayer, and monolayer CuSbS$_2$ in the in-plane directions calculated using GW+BSE. The lowest energy optical transition in each structure is marked with an arrow.}
    \label{fig:dielectric_BSE}
\end{figure}

Having clarified the electronic characteristics of low-dimensional CuSbS$_2$, we complete our analysis by inspecting its optical properties, calculated by solving the BSE~\cite{salpeter1951relativistic} of MBPT. 
Bright excitations below the fundamental band gap occur at 1.20, 0.95, and 0.76 eV for the bulk material, bilayer, and monolayer, respectively, as a signature of excitonic effects (see Figure~\ref{fig:dielectric_BSE} as well as Figure~S2 for the real part and the out-of-plane components). The contributions of the BSE eigenstates to each exciton are represented as a fat band plot in reciprocal space (Figure S3). The contributions between the direct electron-hole pairs are predominately from the bands near the band gap. The strength of the contributions is more in the layered materials which have a reduced band dispersion. The exciton binding energy, calculated as $E_\textrm{b} = E_\textrm{g}^{\textrm{GW}} - E_{\textrm{optical}}$, is 0.20 eV in the bulk where $E_\textrm{g}^{\textrm{GW}}$ is the direct gap from GW and $E_{\textrm{optical}}$ is the lowest energy excitation from BSE. The reduced dimensionality of the system increases the exciton binding energy to 0.35~eV in the bilayer and to 0.54 eV in the monolayer (see Table~\ref{tab:optical}). 
The experimental binding energy is similar to the prediction from theory despite the natural size distribution of platelets in the powder samples~\cite{moosakhani2018solution}. Furthermore, the effects of ligand passivation, defects, or stacking of monolayers into towers could obscure the band gap crossover and optical properties~\cite{moosakhani2018effect}. Experimental detection of the changes in the excitonic peaks or lattice would be improved if the ensemble contains only monolayer platelets. We also note that the possible exciton-like peaks in the layered material may be resolved at lower temperatures which was previously shown for bulk CuSbS$_2$~\cite{birkett2018band}.

\begin{figure}[hbt!]
\includegraphics[width=\textwidth]{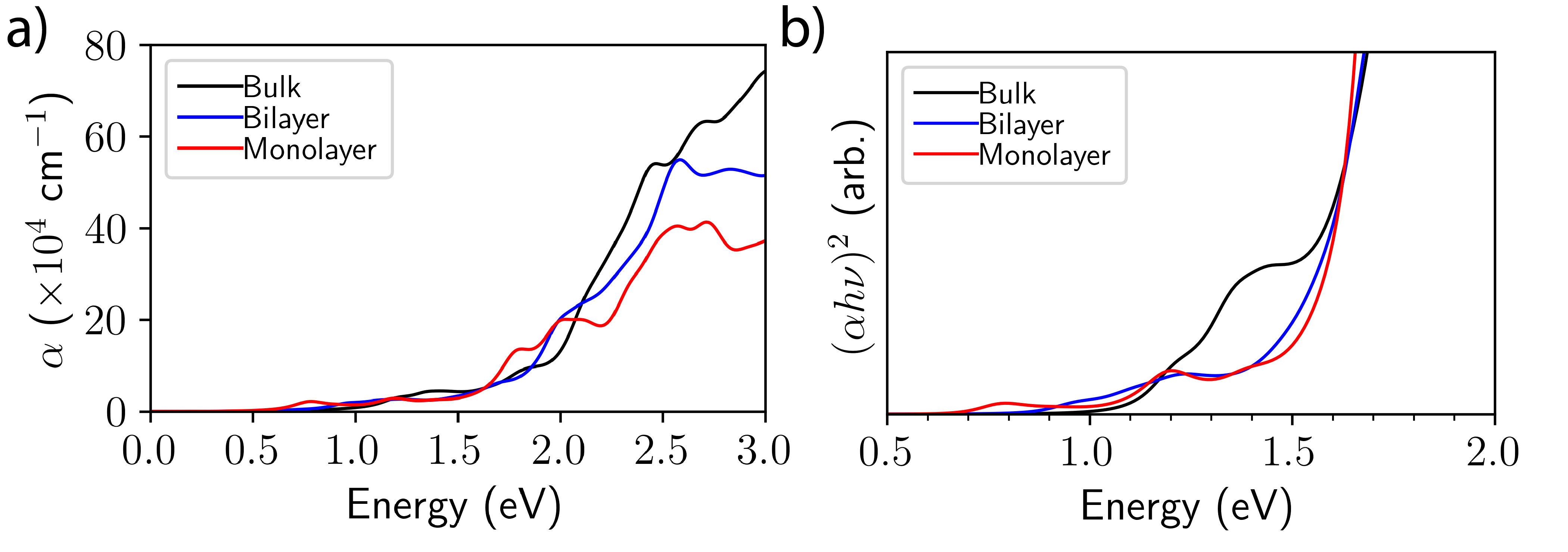}
    \caption{a) Absorption coefficient, $\alpha$, and b) Tauc plot for bulk, bilayer, and monolayer CuSbS$_2$ in the in-plane direction calculated using GW+BSE.}
    \label{fig:tauc_BSE}
\end{figure}

For a closer comparison with experimental results, it is convenient to display the optical properties of each material in form of the absorption coefficient and Tauc plot. The absorption coefficient is $\alpha = 4\pi \kappa / \lambda$, where $\kappa$ is the extinction coefficient and $\lambda$ the wavelength of the incident light. From the plot in Figure~\ref{fig:tauc_BSE}a, we notice that the absorption coefficient increases above $10^5$ cm$^{-1}$ between 1.7 and 1.9 eV regardless of dimensionality. In particular, the Tauc plot in Figure~\ref{fig:tauc_BSE}b shows that \textit{(i)}, despite the spatial confinement in the low dimensional structures, the absorption increases steeply at about 1.5 eV in the bulk, bilayer, and monolayer; \textit{(ii)} weak optical excitations appear below the band gap in all materials, as anticipated in the discussion of the dielectric permittivity (Figure~\ref{fig:dielectric_BSE}). The first finding \textit{(i)} is consistent with recent results from diffuse reflectance spectroscopy on platelets which detected only a small decrease (0.05 eV) in the direct optical gap energy in thin (1 to 5~nm) platelets compared to the thickest platelets (25 to 50~nm)~\cite{moosakhani2018solution}. 
The second finding, \textit{(ii)}, is supported by the presence of weak absorption peaks below the optical gap have been detected in the ellipsometric data of bulk CuSbS$_2$~\cite{birkett2018band} and two-dimensional platelets~\cite{moosakhani2018solution}. The thinnest platelets had a peak at 1.04 eV and energies less than 0.85 eV were beyond the spectral range~\cite{moosakhani2018solution}. 

\section{Conclusions}
In summary, by means of DFT and many body calculations, we rationalized the peculiar behavior of CuSbS$_2$ which exhibits a reduction of the lowest optical transition going from the bulk to mono- and bilayer films. 
In addition to the crossover from indirect to direct-band gap materials, bilayer and monolayer films of CuSbS$_2$ exhibit (slightly) reduced band gap compared to their bulk counterpart due to the formation of localized surface states on the dangling facet. 
On top of this, now in agreement with physical intuition, larger exciton binding energies were found upon increasing quantum confinement, which led to the decrease of the absorption onset going to bulk, to bilayer, and mononlayer, with the latter absorbing radiation in the near-infrared region.
By disclosing the fundamental mechanisms behind the peculiar electronic and optical properties of CuSbS$_2$, this study opens important perspectives for nanoengineering this material and related compounds for optoelectronic and photovoltaic applications.

\section{Supporting Information}
The relaxed lattice constants and out-of-plane dielectric permittivities for each structure.

\section{Acknowledgments}
This work was performed as part of the Academy of Finland RADDESS project 314488 and QTF Centre of Excellence program (project 312298) (KC, TAN). We acknowledge computational resources provided by CSC -- IT Center for Science (Finland) and by the Aalto Science-IT project (Aalto University School of Science). The work was supported in part by the High Performance Computing Centre Stuttgart (HLRS) and HPC-Europa3. C.C. acknowledges financial support from the German Research Foundation, project number 182087777 (CRC 951), from the German Federal Ministry of Education and Research (Professorinnenprogramm III), and from the State of Lower Saxony (Professorinnen für Niedersachsen). 

\bibliography{main_article.bbl}

\clearpage
\section{TOC Graphic}
\begin{figure}
    \centering
    \includegraphics{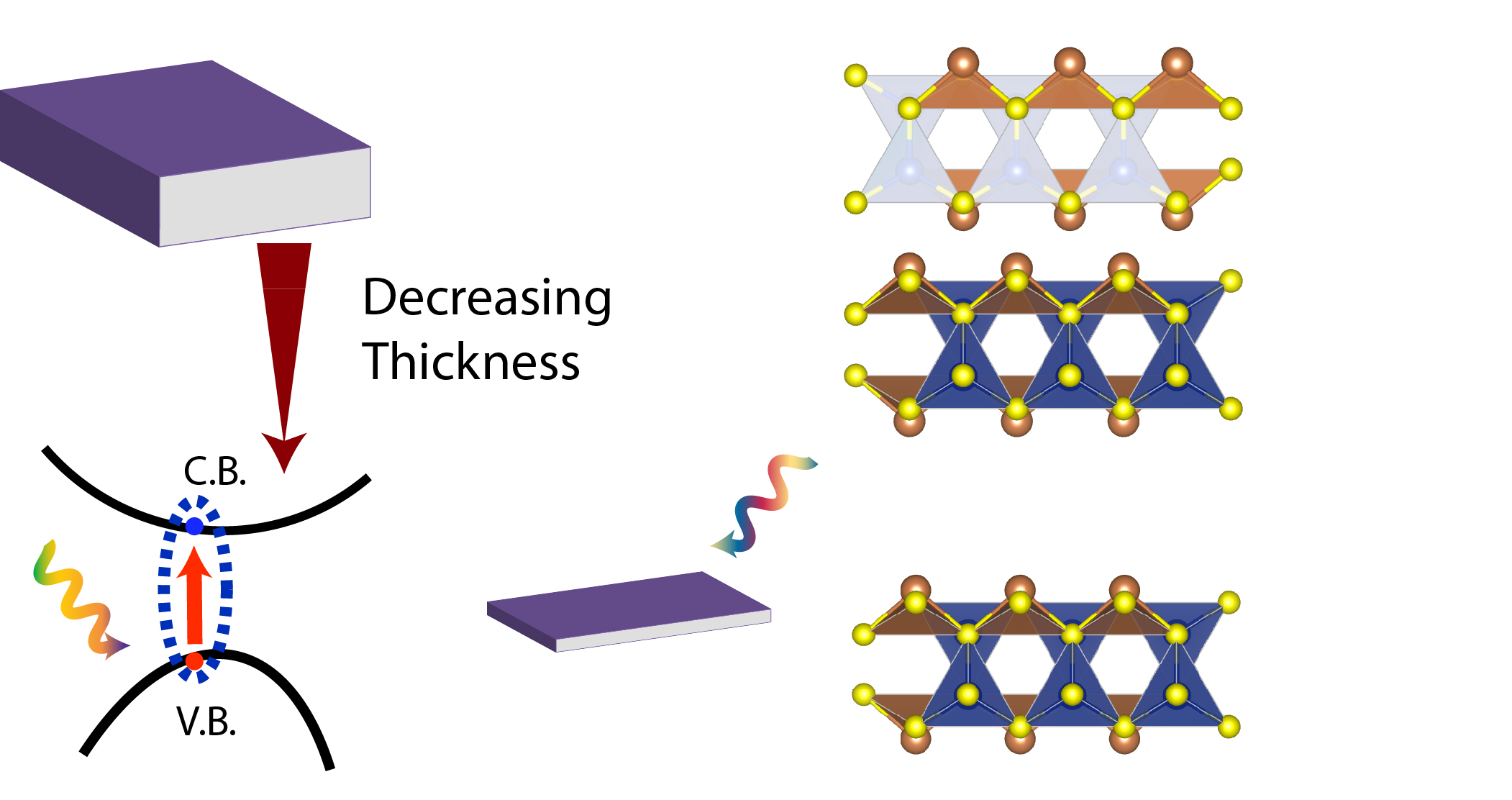}
    \label{fig:TOC}
\end{figure}

\end{document}